\documentclass[reprint, aps,prb,superscriptaddress,showpacs,longbibliography]{revtex4-1}

\usepackage{graphicx,epstopdf,url,times}
\usepackage[colorlinks=true,urlcolor  = blue,citecolor = blue, allcolors=blue]{hyperref}
\usepackage[nameinlink,noabbrev,capitalize]{cleveref}
\usepackage{gensymb}
\usepackage{amsmath}
\usepackage{float}
\usepackage{balance}

\begin{document}

\title{Multimode Directionality in All-Dielectric Metasurfaces}

\author{Yuanqing Yang}
\affiliation{State Key Laboratory of Modern Optical Instrumentation, College of
Optical Science and Engineering, Zhejiang University, Hangzhou 310027, China}

\author{Andrey E. Miroshnichenko}
\affiliation{Nonlinear Physics Center, Research School of Physics and
Engineering, Australian National University, Canberra ACT 2601, Australia}

\author{Sarah V. Kostinski}
\affiliation{Nonlinear Physics Center, Research School of Physics and
Engineering, Australian National University, Canberra ACT 2601, Australia}
\affiliation{Department of Physics, Harvard University, Cambridge, MA 02138,
USA}

\author{Mikhail Odit}
\affiliation{Department of Nanophotonics and Metamaterials, ITMO University,
St.Petersburg 197101, Russia}

\author{Polina Kapitanova}
\affiliation{Department of Nanophotonics and Metamaterials, ITMO University,
St.Petersburg 197101, Russia}

\author{Min Qiu}
\email{minqiu@zju.edu.cn}
\affiliation{State Key Laboratory of Modern Optical Instrumentation, College of
Optical Science and Engineering, Zhejiang University, Hangzhou 310027, China}

\author{Yuri Kivshar}
\email{Yuri.Kivshar@anu.edu.au}
\affiliation{Nonlinear Physics Center, Research School of Physics and
Engineering, Australian National University, Canberra ACT 2601, Australia}
\affiliation{Department of Nanophotonics and Metamaterials, ITMO University,
St.Petersburg 197101, Russia}

\date{\today}

\begin{abstract}
We demonstrate that spectrally diverse multiple magnetic dipole resonances can
be excited
in all-dielectric structures lacking rotational symmetry, in contrast to 
conventionally used spheres, disks or spheroids. Such multiple magnetic
resonances arise from hybrid Mie-Fabry-P\'{e}rot modes, and can constructively
interfere with induced electric dipole moments, thereby leading to
novel multi-frequency unidirectional scattering. Here we focus on elongated
dielectric
nanobars, whose magnetic resonances can be spectrally tuned by their aspect
ratios. Based on our theoretical results, we suggest all-dielectric multimode
metasurfaces and verify them in proof-of-principle microwave experiments. We
also believe that the demonstrated property of multimode directionality is
largely responsible for the best efficiency of all-dielectric metasurfaces that
were
recently shown to operate across multiple telecom bands.
\end{abstract}


\maketitle

\section{Introduction}

Modern nanophotonics aims to efficiently manipulate light at the nanoscale, with
applications ranging from near-field microscopy and integrated optoelectronics
to biomedical science \cite{novotny2012principles}. Recent decades have
witnessed a growing research interest in the study of \emph{plasmonic
nanoparticles} made of gold or silver, recognized for their outstanding ability
to squeeze light into subwavelength volumes via surface plasmon resonances. The
resonant optical plasmonic modes supported by metallic structures endow them
with an ability to manipulate light at subwavelength scales. These optical
resonances are highly dependent on the choice of the structure's material and
geometry, allowing for further manipulations. Various types of photonic devices
based on plasmonic nanoparticles have thus been demonstrated
\cite{*{novotny2012principles,schuller2010plasmonics,novotny2011antennas,kinkhabwala2009large,curto2010unidirectional}}.
However, their overall functionalities and performance are severely affected by
high intrinsic losses in metals. When larger amounts of metals are involved in
complex plasmonic structures such as metamaterials or metadevices
\cite{*{gramotnev2010plasmonics,kildishev2013planar,zheludev2012metamaterials}},
the loss problem is exacerbated and hinders their scalability for practical use.

Whereas new materials with improved plasmonic properties have been proposed,
there has also been a growing realization that the optical resonances of
high-index resonant dielectric structures can facilitate light manipulation
below the free-space diffraction limit with very low losses
\cite{jahani2016all,kuznetsov2016optically,evlyukhin2012demonstration,kuznetsov2012magnetic,slobozhanyuk2015subwavelength,miroshnichenko2015nonradiating,staude2013tailoring,luk2015optimum,ginn2012realizing,sikdar2015optically,vynck2009all,fan2014optical,yang2015controlling,ee2015shape,tian2016tailoring,fu2013directional,krasnok2011huygens,liu2014optical}.  In contrast to plasmonic nanoparticles that are dominated by electric resonances, 
high-refractive-index dielectric nanoparticles have proven to support both \emph{electric and magnetic} 
Mie-type dipole and multipole resonances, opening up new possibilities for designer
photonic metadevices
\cite{*{jahani2016all,kuznetsov2016optically,evlyukhin2012demonstration,kuznetsov2012magnetic,slobozhanyuk2015subwavelength,miroshnichenko2015nonradiating,staude2013tailoring,luk2015optimum,ginn2012realizing,sikdar2015optically}}.
For example, by using an isolated magnetic dipole Mie resonance, a magnetic
mirror can be realized \cite{liu2014optical}. While if we use a magnetic dipole
that is spectrally overlapped with an electric dipole, these two dipole modes
can satisfy the first Kerker condition \cite{kerker1983electromagnetic} and
constructively interfere with each other, leading to directional scattering and
the realization of transparent Huygens' metasurfaces
\cite{jahani2016all,kuznetsov2016optically}. Therefore, how to fully exploit these
intriguing optically-induced electric and magnetic resonances becomes
extremely crucial for realizing and functionalizing dielectric metasurfaces.

\begin{figure}[h]
\includegraphics[width=8.6cm]{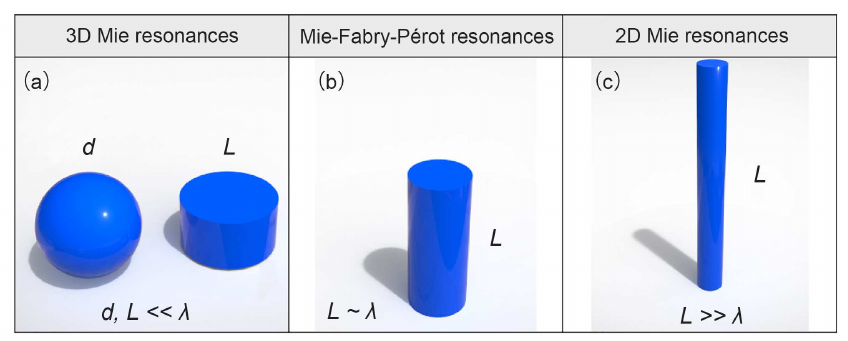}
\caption{\label{1}  Classes of all-dielectric meta-atoms:
(a) Sphere and nanodisk with high refractive index described by the
three-dimensional Mie scattering theory. Characteristic dimensions ($d$ and $L$)
are much smaller than the free-space wavelength $\lambda$. (b) Finite-size
nanorod ($L \sim \lambda$) with a high aspect ratio supporting the hybrid
Mie-Fabry-P\'{e}rot as described in this work. (c) Long nanorod ($L \gg
\lambda$) described by the two-dimensional Mie scattering theory.}
\end{figure}

However, in all studied dielectric resonant structures presented so far, the
geometry of dielectric nanoparticles is considered to be close to either spheres
\cite{evlyukhin2012demonstration,kuznetsov2012magnetic,slobozhanyuk2015subwavelength},
spheroids/disks
\cite{luk2015optimum,miroshnichenko2015nonradiating,staude2013tailoring}, cubes
\cite{ginn2012realizing,sikdar2015optically}, or long rod
\cite{vynck2009all,fan2014optical} [see Figs. 1(a,c)], so the exact Mie
solutions of the two- (2D) and three-dimensional (3D) scattering problems can be
applied to analyze the scattering by such isotropic or symmetric nanostructures. 
These symmetric structures, as verified by Mie theory and associated multiple
expansions, can support a series of different resonances, with
first-order Mie resonance usually a \emph{single} magnetic dipole mode, the
second-order a \emph{single} electric dipole and subsequent higher-order
electric and magnetic multipoles. By contrast, if we consider dielectric
nanoparticles with broken rotational symmetry such as finite-size nanobars [see Fig.
1(b)], as we will show in the following, such asymmetric meta-atoms will not
only introduce new physics into the classical Mie scattering problem but can
also bring novel functionality to all-dielectric structures and metasurfaces.  

In this paper, we focus on silicon nanobars with \emph{a large aspect ratio} and 
demonstrate that such elongated nanostructures can support hybrid Mie-Fabry-P\'{e}rot modes 
associated with \emph{multiple} magnetic dipole resonances. These intriguing modes 
arise from the combination of conventional magnetic dipole modes 
excited in the transverse direction ({\em Mie resonances}) and  the standing waves 
excited in the longitudinal direction  ({\em Fabry-P\'{e}rot cavity modes}). Moreover, just like single 
magnetic dipoles, such multiple magnetic dipole modes can also constructively interfere with induced 
electric dipoles, thereby leading to multi-frequency directional scattering, 
characterized by multiple Kerker conditions. Based on our theoretical results, we further demonstrate 
novel all-dielectric Huygens' metasurfaces with spectrally diverse directionality
verified in proof-of-principle microwave experiments. Due to the existence of
multiple magnetic dipoles, such metasurfaces can work efficiently in both
reflection and transmission modes and also achieve all four quadrants of electromagnetic responses:
$\epsilon > 0, \mu > 0$; $\epsilon < 0, \mu > 0$; $\epsilon > 0, \mu < 0$; $\epsilon < 0, \mu > 0$, 
where $\epsilon$ and $\mu$ are electric permittivity and magnetic permeability, respectively. 
It is also worth noting that whereas there are some 
recent efforts on metasurfaces using dielectric building blocks with broken rotational 
symmetry \cite{devlin2016broadband, khorasaninejad2016super, wang2014generation}, 
most designs do not directly rely on the resonances of single elements \cite{kuznetsov2016optically, devlin2016broadband, khorasaninejad2016super} 
and only fundamental electric and magnetic dipole modes have been studied \cite{wang2014generation}.
Finally, we also argue that that the operation of the 
recently demonstrated broadband all-dielectric metasurfaces \cite{kruk2016invited} is 
largely due to the multiple magnetic multipole modes of the constituent elements in the 
form of tall dielectric rods, allowing to achieve destructive interference in reflection 
over a large spectra bandwidth. Our findings are expected to provide a new methodology 
to design broadband and multifunctional all-dielectric metadevices.

\section{Scattering and multipole decomposition}

\begin{figure}
\includegraphics[width=8.6cm]{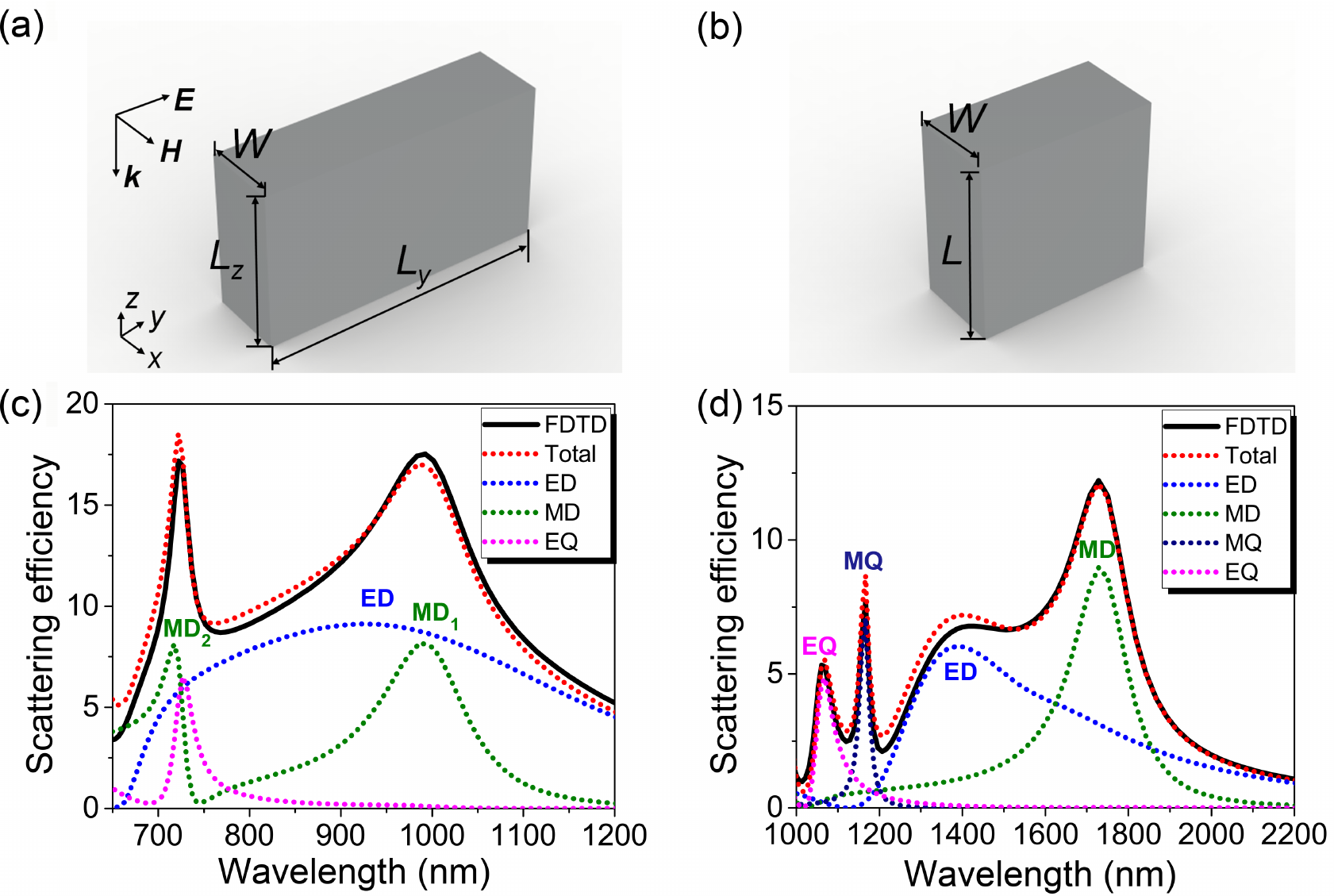}
\caption{\label{2}(a, b) Schematic diagrams of (a) an anisotropic ($W$ = 110 nm,
${L_y}$ = 400 nm and ${L_z}$ = 220 nm) and (b) a symmetric (${W}$ = ${L}$ 
= 400 nm) silicon nanobar. (c, d) Simulated scattering spectra
(solid black line) and calculated multipole decompositions (total contributions:
dotted red line, ED: dotted blue line, MD: dotted green line, EQ: dotted magenta
line) of (c) the anisotropic and (d) symmetric nanobar, respectively.}
\end{figure}

The schematic of a designed silicon anisotropic nanobar is shown in Fig. 2(a).
The geometric parameters are all different in three dimensions with $W$ = 110
nm, $L_z$ = 220 nm and $L_y$ = 400 nm. For comparison, we also introduce a
symmetric silicon nanobar with  ${L_y = L_z = L= W}$ = 400 nm, as depicted in Fig. 2(b). 
Here we use 3D finite-difference time-domain (FDTD) simulations \cite{FDTDSolutions} 
and the Cartesian multipole analysis [see \cref{Appendix:A}]
to calculate the response of the structures and identify the contributions from
each multipole moments. The optical constants of silicon is taken from Palik's handbook
\cite{palik1998handbook} while the surrounding media is assumed to be air. The
structures are illuminated by a normally incident plane wave with electric field
along $y$ direction.

Fig. 2(c) and 2(d) represent the calculated scattering efficiency spectra and
decomposed multipole contributions. The scattering efficiency $Q_{\textnormal{eff}}$ is
defined as $Q_{\textnormal{eff}} = Q_{\textnormal{sca}}/Q_{\textnormal{geo}}$, 
where $Q_{\textnormal{sca}}$ and $Q_{\textnormal{geo}}$ are 
scattering and geometrical cross sections of the particle, respectively. Here in
our case, $Q_{\textnormal{geo}} = W \times \ L_y$. 
For multipole expansions, we only consider 
the first four terms, namely, electric dipole (ED), magnetic dipole (MD) and
electric quadrupole (EQ) and magnetic quadrupole (MQ) modes. 
The scattering spectra obtained from the FDTD simulations (solid black line) 
and the multipole expansions (dotted red line) are in a good agreement with each other, 
indicating that higher-order multipoles are negligible in our case. At first glance, 
both scattering spectra of the nanobars have similar optical responses with two pronounced 
maxima [cf. dotted black curves in Figs. 2(c,d)]. However, through the multipole expansion, 
we reveal that the underlying contributions of each multipole moments to these
peaks are \emph{completely different}. For the symmetric nanobar, the
peaks are attributed to the separated MD and ED resonant modes, as has been
reported in many previous studies on all-dielectric spheres, disks, or cubes. By
contrast, the first peak in the scattering spectrum of the anisotropic nanobar
shows a resonance overlap of MD and ED, while the second peak arises from the
second maximum in the magnetic dipole contribution, implying the existence of a
second-mode magnetic dipole (MD$_2$), which has never been discussed or
demonstrated before. We would also like to note that this MD$_2$ mode
is essentially different from conventional MQ mode, which will be shown
in the following section.

\section{Multi-frequency directional scattering}

\begin{figure*}
\onecolumngrid
\includegraphics[width=17.2cm]{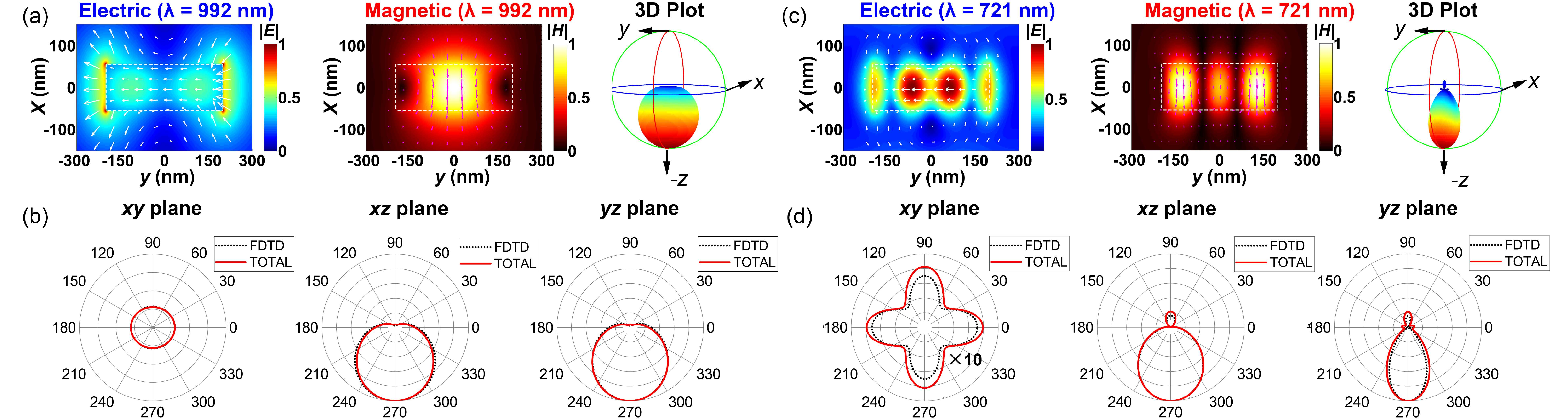}
\caption{\label{3}(a, c) Near-field distributions in the middle cut plane ($z$ =
0) and 3D far-field scattering patterns of the nanobar at (a) $\lambda$ = 992 nm
and (c) $\lambda$ = 721 nm, respectively. The colors represent normalized
amplitudes of the electric and magnetic fields, and arrows show the field
vectors. (b, d) Simulated far-field scattering patterns (dotted black line) and
calculated multipole radiation patterns (solid red line) at (b) $\lambda$ = 992
nm and (d) $\lambda$ = 721 nm, respectively. The patterns are normalized to the
maximum scattering intensity in the far field.}
\twocolumngrid
\end{figure*}

To further illustrate the properties of the isotropic nanobar and especially the
MD$_2$ mode, in Fig. 3 we plot the near- and far-field distributions at two peak
wavelengths ($\lambda$ = 992 nm and $\lambda$ = 721 nm). For $\lambda$ = 992 nm,
the induced ED (parallel to the incident polarization, $p_{y1}$) and MD
($m_{x1}$) dominate the near-field profiles with very close amplitudes
(${|p_{y1}| = 1.03 \times |m_{x1}|/c}$, where $c$ is the speed of light in
vacuum) and a moderate phase difference (${\Delta \phi \sim 23^{\circ}}$),
making them approximately satisfy the first Kerker condition
\cite{kerker1983electromagnetic} and thus resulting in unidirectional forward
scattering along $z$ direction, as shown in Figs. 3(a,b). The simulated
scattering patterns (dotted black lines) are also in an excellent accordance
with the calculated radiation patterns from decomposed multipoles (solid red
lines).

Fig. 3(c) shows contrasting field distributions at $\lambda$ = 721 nm. We
observe that standing wave patterns appear in both electric and magnetic fields,
providing valuable insights into the nature of the MD$_2$ mode. The electric
field is the superposition of a standing wave $E_z$ and an induced ED mode
($p_{y2}$) in $y$ direction, whereas the magnetic field is the consequence of a
standing wave $H_x$ along with an induced MD mode in $x$ direction as well,
leading to the appearance of the hybrid Mie-Fabry-P\'{e}rot mode MD$_2$ 
[see \cref{Appendix:B} for theoretical standing wave decompositions]. In spite of the
standing wave pattern or fluctuations in the magnetic field distribution, the
MD$_2$ mode still has a net magnetic dipole moment ($m_{x2}$) in $-x$ direction,
just like the fundamental MD mode that we call now MD$_1$ mode. Interestingly,
this magnetic dipole moment can also nearly satisfy the first Kerker condition
with the electric dipole (${|p_{y2}| = 0.98 \times |m_{x2}|/c}, {\Delta \phi
\sim 13^{\circ}}$), thereby offering the novel behavior of multimode
(multi-frequency) unidirectional scattering [see \cref{Appendix:C} for theoretical explanations]. 
This unique property is clearly shown in Fig. 3(d). We can find good agreement between 
the simulated and calculated angular patterns. Meanwhile, 
we should remember about the existence of the EQ mode. Although it brings about 
small undesired backscattering, it also substantially narrows the scattering pattern and boosts 
the directivity. A front-to-back power ratio higher than 9 thus could be obtained in this case.

Besides the two well defined maxima in the scattering spectrum, there is also a
noticeable dip around $\lambda$ = 767 nm [see in Fig. 2(c)], accompanied by a
minimum near zero in the MD contribution, indicating that the contribution of
the MD mode to the far field almost vanishes. This dip can be attributed to 
the cancellation of the induced magnetic dipoles which have opposite directions 
in the anti-nodes of the standing-wave pattern, mimicking a magnetic `dark mode'. 
Specifically, the amplitude of the net magnetic dipole moment at $\lambda$ = 767 nm 
is only $\sim$1/5 of that of the electric dipole moment, corresponding to $\sim$1/25 in
the far-field contributions.

\begin{figure}[b]
\includegraphics[width = 8.6cm]{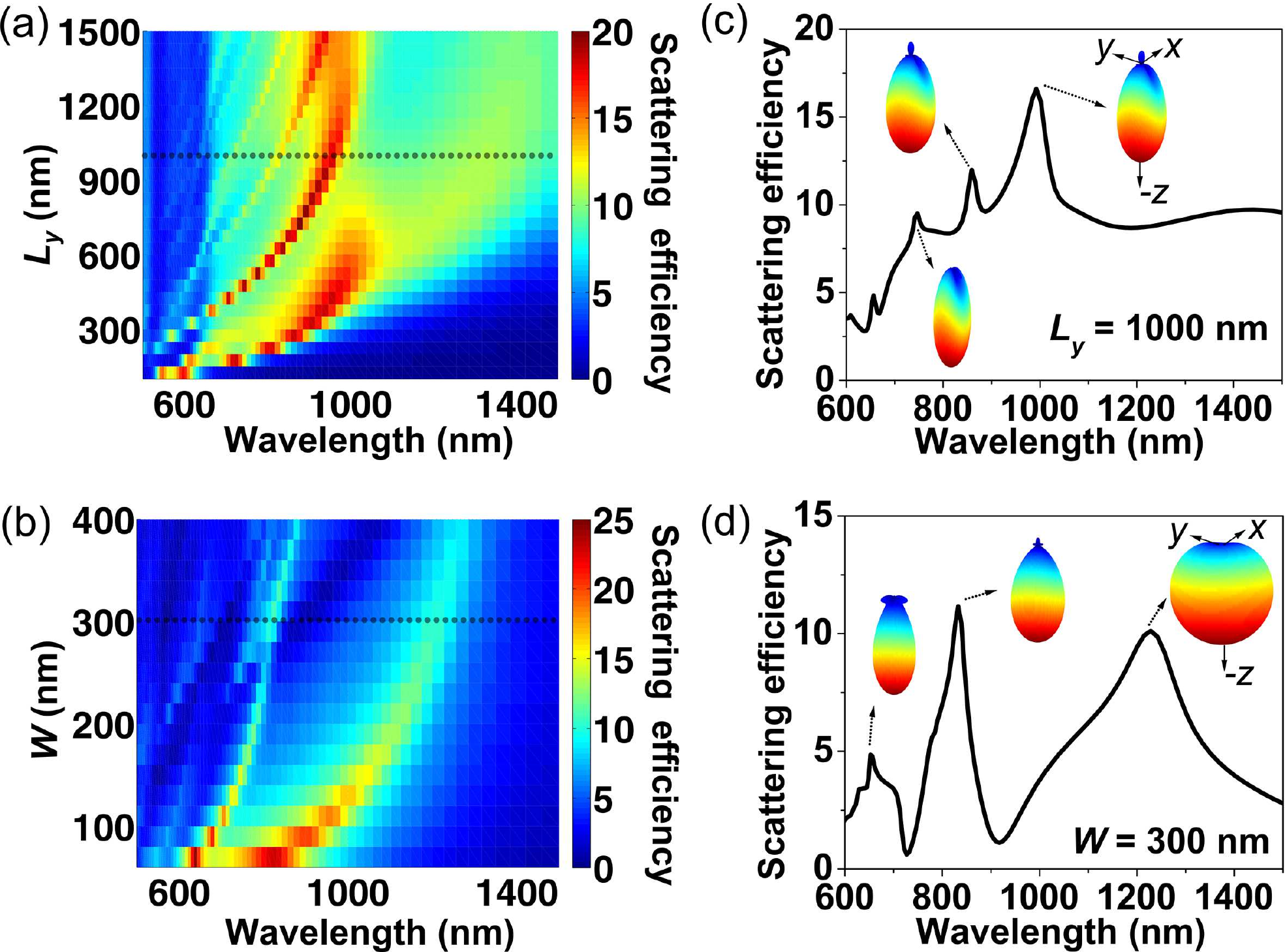}
\caption{\label{4}Scattering efficiency spectra as a function of geometric
parameters (a) $L_y$ with fixed $W$ =110 nm and $L_z$ = 220 nm, and (b) $W$ with
fixed $L_z$ = 220 nm and $L_y$ = 400 nm. (c, d) Scattering spectrum for a
nanobar with dimensions marked by the dashed lines in (a) and (b)
correspondingly. The insets show the far-field unidirectional scattering
patterns at different resonance wavelengths.}
\end{figure}

\begin{figure*}[t]
\onecolumngrid
\includegraphics[width = 17.2cm]{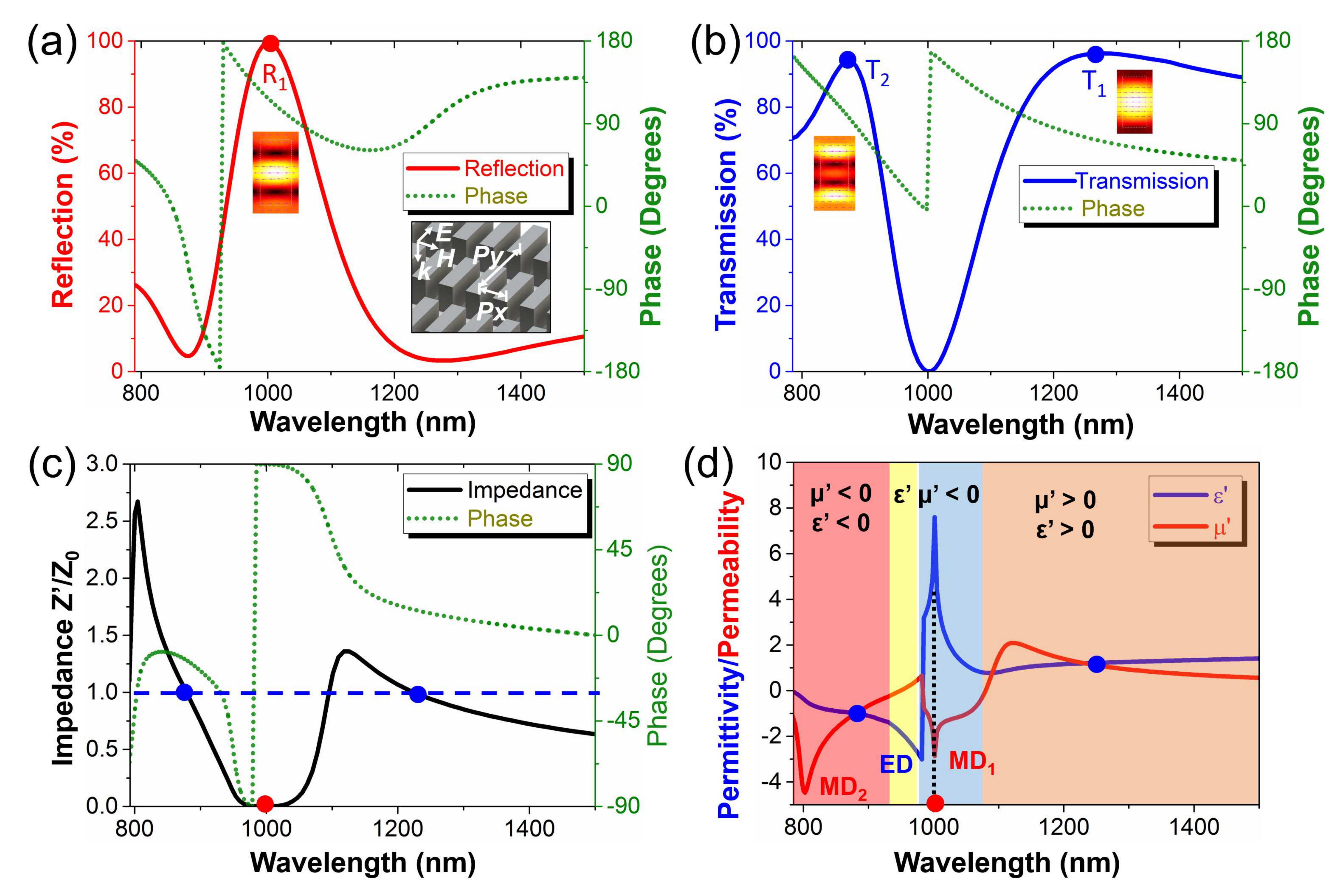}
\caption{\label{5}(a, b) Reflection and transmission spectra of the 
metasurface composed of the anisotropic nanobars shown in FIG. 2a. The periodicities 
in $x$ and $y$ directions are 160 nm and 500 nm, respectively. Insets diagram the configuration
and corresponding near-field magnetic distributions in $xy$ plane at R$_{1}$, T$_{1}$, T$_{2}$ 
peak wavelengths. (c) Calculated impedance of the metasurface. Solid black line is the real value $Z'$ 
of the impedance $Z$, which corresponds predominately to radiation resistance. 
Green dotted line is the impedance phase. The blue dashed line indicates the impedance matching 
condition $Z' = Z_{0}$ (d) Effective permittivity (blue) and permeability (red) of the metasurface obtained 
using S-parameter retrieval.$\epsilon'$ and $\mu'$ denote the real parts of $\epsilon$ and $\mu$.}
\twocolumngrid
\end{figure*}

Since the MD$_2$ mode arises from a magnetic standing wave pattern, one can
intuitively expect a strong dependence of geometric parameters on the mode
characteristics and further contributions to the scattering properties. In Fig.
4(a) and Fig. 4(c), we use two-dimensional color maps to show the impact of the
geometric parameters $W$ and $L_y$ on the scattering spectra. With increasing
length $L_y$ and width $W$, we can see evident red-shifts and the newly emerged
higher-order Mie resonances. These red-shifts and new Mie resonances, along with
the Fabry-P\'{e}rot resonances, can further lead to other multimode ED and MD
besides the MD$_2$ mode. For instance, the scattering spectra for nanobars with
$L_y$ = 1000 nm and $W$ = 300 nm, marked by the dashed black lines in the 2D
color maps, show a fascinating property of triple-wavelength unidirectional
scattering supported by an individual nanobar [Figs. 4(b,d)]. This is exactly
due to the interferences of the multimode MD and ED as well as other multipole
moments excited inside the nanobars with increasing geometric parameters,
accompanied by increasing-order Fabry-P\'{e}rot modes. In particular, it can be
seen that, the increase in $L_y$ results in higher-mode MD while the increase in
W brings about higher-mode ED [see \cref{Appendix:D}].

\section{Multimode metasurfaces}

Since the presented individual nanobar have proven to support multifrequency 
directional scattering, we expect that a mteasurface composed of such nanobars 
can also have a multimode response. In Fig. 5 we plot reflection and transmission
full spectra (intensity and phase) of such a metasurface. The inset diagrams 
the metasurface with  $P_{x}$ = 160 nm and $P_{y}$ = 500 nm  
(periodicities in $x$ and $y$ directions) on a glass substrate ($n_{\textnormal{glass}}$ = 1.5). 
One reflection peak R$_1$ and two transmission peaks T$_1$ and T$_2$ can be
seen in the plots, indicating that our metasurface can function as either a perfect mirror or
a transparent film at different wavelengths. At transmission peak T$_1$, the fundamental 
electric and magnetic dipole moments (ED$_1$ and MD$_1$) constructively interfere with 
each other and lead to the high transmission. While at the high reflection peak R$_1$, 
a standing wave pattern appears and the magnetic dipole moment has an opposite 
direction to that at T$_1$. With the electric field kept in the same direction, this will lead 
to a reversal in the direction of power flow, i.e. changing from high transmission to 
high reflection. By contrast, at second transmission peak T$_2$, the hybrid magnetic 
dipole moment once again has the same direction as that in T$_1$, thereby resulting in 
a second high transmission peak. This phase-flipping phenomenon of the magnetic dipoles 
and associated multimode high transmission are directly due to the emergence of 
MD$_2$ modes. Moreover, these multiple resonant modes also enable both 
reflected and transmitted light to experience significant phase changes with 
maintained high efficiency. The abrupt phase changes arising from the resonances 
can be easily tuned by varying the sizes of the nanostructures, which can be further 
used in the implementation of perfect reflectors, magnetic mirrors or gradient metasurfaces \cite{jahani2016all, kuznetsov2016optically}.
Compared to previous studies where metasurfaces can only work in reflection or 
transmission modes, our metasurface makes it possible to control both reflected 
and transmitted light, which can remarkably extend the functionality of metasurfaces. 

FIG. 5 (c) shows the calculated impedance of the metasurface. The two transmission 
peaks T$_1$ and T$_2$ correspond well to the impedance-matched points 
while reflection peak R$_1$ corresponds to a largely mismatched point where 
the wave impedance becomes predominately imaginary. A striking flip of the impedance 
phase also occurs around 990 nm from +90\degree to -90\degree, indicating the 
metasurface switching fast from a magnetic conductor to an electric conductor \cite{ginn2012realizing}.

To better understand the optical response of the metasurface, we also adopted a 
standard S-parameter retrieval method \cite{smith2002determination} to compute the 
effective permittivity and permeability, as shown in Fig. 5 (d). Two notable magnetic 
resonances and one electric resonance could be observed. Combing the corresponding 
near-field distributions, it is easy to verify the existences of the ED, MD$_1$ and 
MD$_2$ mode induced in the metasurface. The spectral positions of these modes 
are different from those induced in the individual nanobar because of the substrate effect 
and the mutual interaction. Two intersections between the plots of permittivity and permeability 
indicate the impedance matched points and the fulfillment of the Kerker condition. 
The first transmission T1 appears at the tails of the fundamental ED and MD$_1$ resonances, 
showing an “off-resonant” directionality. In this region ($\lambda> 1080$  nm), the permittivity 
and permeability of the metasurfaces are both above zero, which means the overall response 
of the metasurface is similar to conventional dielectric materials. However, for shorter 
wavelengths, the electric and magnetic resonances lead to distinct phenomena. 
The MD$_1$ mode makes the metasurface function as a magnetic mirror which has a negative 
permeability ($\mu < 0$) while the ED mode enables the metasurface to function as an electric 
mirror with a negative permittivity ($\epsilon < 0$). More interestingly, these two contrasting 
behaviors can be switched to each other very fast since the ED and MD$_1$ modes are spectrally 
very close to each other. This is also in good accordance with the impedance phase flip 
occurring at 980 nm. Another fascinating feature of the metasurface is its negative refractive index 
($\epsilon < 0$, $\mu< 0$ ) attributed to the MD$_2$ and the ED modes for $\lambda < 950$ nm. 
In this region, the constructive interference of the MD$_2$ and ED modes happens in both of their 
resonance regimes, resulting in an efficient Huygens’ source with negative permittivity 
and permeability. Therefore, our metasurface can support all four quadrants of 
possible optical responses, which can bring various unexplored possibilities and functionalities.

\begin{figure}[t]
\includegraphics[width = 8.6cm]{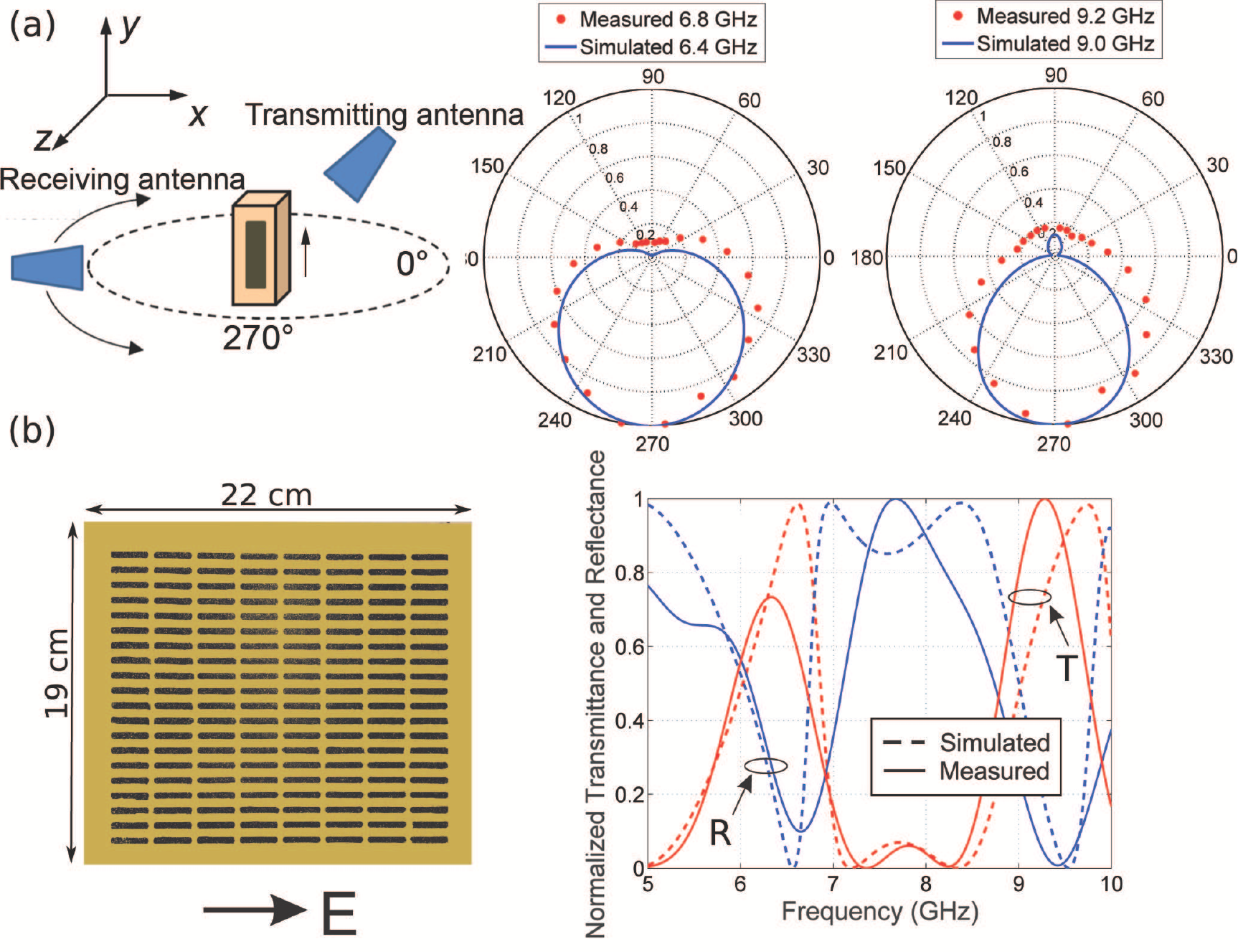}
\caption{\label{6}(a) Dielectric bar scatter. Left: a sketch of the
experimental setup to measure the radiation pattern of the single scatter with
the dimensions $W$ = 0.5 cm, $L_y$ = 1.8 cm, $L_z$ = 1.5 cm. Right:
experimentally measured (red dots) and CST numerically simulated (solid curves)
radiation patterns for the Kerker conditions. (b) Dielectric metasurface. Left:
photograph of the fabricated multimode Huygens' metasurface composed of
anisotropic dielectric bars with the dimensions $W$ = 0.5 cm, $L_y$ = 1.8 cm,
$L_z$ = 1.5 cm placed with the periods $P_x$ = 0.9 cm and $P_y$ = 2.7 cm.
Right:Experimentally measured (solid curves) and CST numerically simulated
(dashed
curves) reflection and transmission spectra magnitudes of the multimode
Huygens'metasurface.}
\end{figure}

To verify the proposed concept experimentally, the silicon nanobars are 
scaled up to the microwave frequency range. Here we employ full-scale numerical
simulations \cite{CST} to optimize bar scatterers and use Eccostock HiK ceramic
powder (permittivity $\varepsilon$ = 10 and loss tangent tan $\theta$ = 0.0007)
as the high-index dielectric material to mimic silicon nanobars in the microwave
region.

First, we study experimentally the scattering from a single bar scatter in an
anechoic chamber. The experimental setup is sketched in Fig. 6(a). To perform a
plane wave excitation and to receive the scattered signal, we utilized a pair
of identical rectangular linearly polarized wideband horn antennas (operational
range 1\textendash18 GHz) that were connected to the ports of a Vector Network
Analyzer (Agilent E8362C). The polarization is along $y$ direction. The
transmitting antenna and the single scatter have been fixed, whereas the
receiving antenna was moving around the scatter in $xz$ plane. The scattering
cross-section patterns measured in $xz$ plane at two distinct frequencies $f$ =
6.8 GHz and $f$ = 9.2 GHz are plotted in Fig. 6(a) and they are compared with
the results of numerical simulations. We find the best agreement for slightly
shifted frequencies $f$ = 6.4 GHz and $f$ = 9.0 GHz, and the difference between
of the measured Mie resonant frequencies and simulated resonances can be
explained by the tolerance of the antenna prototype fabrication. These results
clearly demonstrate the multifrequency directional scattering supported by a single
dielectric bar scatter.

Next, we consider all-dielectric metasurfaces composed of the elongated
anisotropic bars. A photograph of the experimental metasurface prototype is
shown in Fig. 6(b). The transmission and reflection spectra of the metasurface
have been investigated both numerically and experimentally. We observe that the
metasurface exhibits an expected multimode response with two pronounced maxima
in the transmission coefficient (at the frequencies around 6.5 GHz and 9.5 GHz)
and one well-defined peak in the reflection coefficient (in the frequency band
7.5\textendash8 GHz), as predicted numerically for the optical frequency range. 
The slight disagreement between the measured and simulated results in the positions 
of frequencies in the transmission/reflection maxima and minima can be explained
by the tolerance of the sample fabrication. The mismatching in the magnitudes of
transmission/reflection coefficients is due to a deviation of permittivity in
each particular unit cell caused by different density of ceramic powder.

\section{Conclusion}
We have presented the novel all-dielectric metasurfaces with 
multimode directionality. Such metasurfaces can support all four possible quadrants 
of electromagnetic responses and can also work efficiently with either high reflection 
or high transmission, which may find many applications and largely extends 
the possibilities of planar optics. We have also demonstrated that this unique
multimode property originates from the hybrid Mie-Fabry-P\'{e}rot modes
supported by high-index dielectric structures with large aspect ratios. The
revealed hybrid modes and associated multiple magnetic dipole resonances also
open an universally new way for engineering the properties of resonant
nanostructures and metamaterials.

We also believe that the phenomenon of multimode magnetic dipole moments is
responsible for the best efficiency of broadband all-dielectric metasurfaces
based on the generalized Huygens principle. Indeed, the superposition of the
scattering contributions from several electric and magnetic multipole modes of
the constituent metaatoms allows to achieve destructive interference in
reflection over a large spectral bandwidth, demonstrating reflectionless
half-wave plates, quarter-wave plates, and vector beam q-plates that can operate
across multiple telecom bands with $\sim$ 99 $\%$polarization conversion
efficiency \cite{kruk2016invited}.

\appendix

\section{Multipole decomposition}\label{Appendix:A}
We employed the Cartesian multipole expansion technique \cite{miroshnichenko2015nonradiating, evlyukhin2013multipole} 
to analyze different multipole modes inside the nanobars. The multipoles are 
calculated through the light-induced polarization $\textbf{P} = \epsilon_0(\epsilon_r-1)\textbf{E}$, 
where $\epsilon_0$ and $\epsilon_r$ are the vacuum permittivity and relative 
permittivity of the nanobar, respectively. We can write \textbf{P} as:
\begin{equation}
\mathbf{P(r)}=\int\mathbf{P(r')}\delta(\mathbf{r}-\mathbf{r'})d\mathbf{r'},
\end{equation}
and then expand the delta function in a Talyor series with respect to \textbf{r'} 
around nanobar's center (origin point \textbf{r$_0$}). Then we can get:
\begin{align}
\mathbf{P(r)} & \simeq \mathbf{p}\delta(\mathbf{r}) + 
\frac{i}{\omega} [\nabla\times\mathbf{m}\delta(\mathbf{r})]
- \frac{1}{6}\hat{Q}\nabla\delta(\mathbf{r})  \nonumber \\
&- \frac{i}{2\omega}[\nabla \times \hat{M} \nabla \delta(\mathbf{r})],
\end{align}
where the multipole moments (electric dipole \textbf{p}, magnetic dipole \textbf{m}, 
electric quadrupole tensor $ \hat{Q} $ and magnetic quadrupole tensor $ \hat{M} $) 
are defined as:
\begin{align}
\mathbf{p} & = \int \mathbf{P(r')} d\mathbf{r'}, \\
\mathbf{m} & = - \frac{i \omega}{2} \int [\mathbf{r'} \times \mathbf{P(r')}] d\mathbf{r'},\\
\hat{Q} & = 3 \int{\mathbf{r'P(r') + P(r')r'} - \frac{2}{3}[\mathbf{r'}  \cdot \mathbf{P(r')}] \hat{U}} d\mathbf{r'} \\
\hat{M} & = \frac{\omega}{3i} \int \{ [\mathbf{r' \times P(r')}]\mathbf{r'} + \mathbf{r' [r' \times P(r')]}\} d\mathbf{r'},
\end{align}
with $ \omega $ is the angular frequency and $ \hat{U} $ is the 3 $\times$ 3 unit tensor. The scattered far-field electric field thus can be calculated by:
\begin{align}
\mathbf{E}_{sca}( \mathbf{r}) & \simeq \frac{k_0^2}{4\pi\epsilon_0} \frac{e^{ik_0r}}{r}  \Big\{ \mathbf{[n \times [p \times n]]} + \frac{1}{c}[\mathbf{m \times n}] \nonumber \\
& + \frac{ik_0}{6}[\mathbf{n \times}[\mathbf{n} \times \hat{Q} \mathbf{n}]] + \frac{ik_0}{2c}[\mathbf{n} \times (\hat{M} \mathbf{n})] \Big\},
\end{align}
in which $ r= |\mathbf{r}|$, $ \mathbf{n} $ is the unit vector directed along \textbf{r}, 
$k_0$ is the wave number and $c$ is the speed of light in a vacuum. The total radiation 
power $P_{sca}$ of the multipoles is:
\begin{align}
P_{sca} & \simeq \frac{c^2k_0^4Z_0}{12\pi}|\mathbf{p}|^2 + \frac{k_0^4Z_0}{12\pi} |\mathbf{m}|^2 + \frac{c^2k_0^6Z_0}{1440\pi} \sum|Q_{\alpha \beta}|^2 \nonumber \\
& +\frac{k_0^6Z_0}{160\pi} \sum |M_{\alpha\beta}|^2,
\end{align}
where $Z_0$ is the vacuum wave impedance and $ \alpha, \beta \equiv x, y, z $ denote Cartesian components.

\section{Field decomposition of a dielectric resonator: theory vs simulations}\label{Appendix:B}

\begin{figure}[t]
\includegraphics[width = 8.6cm]{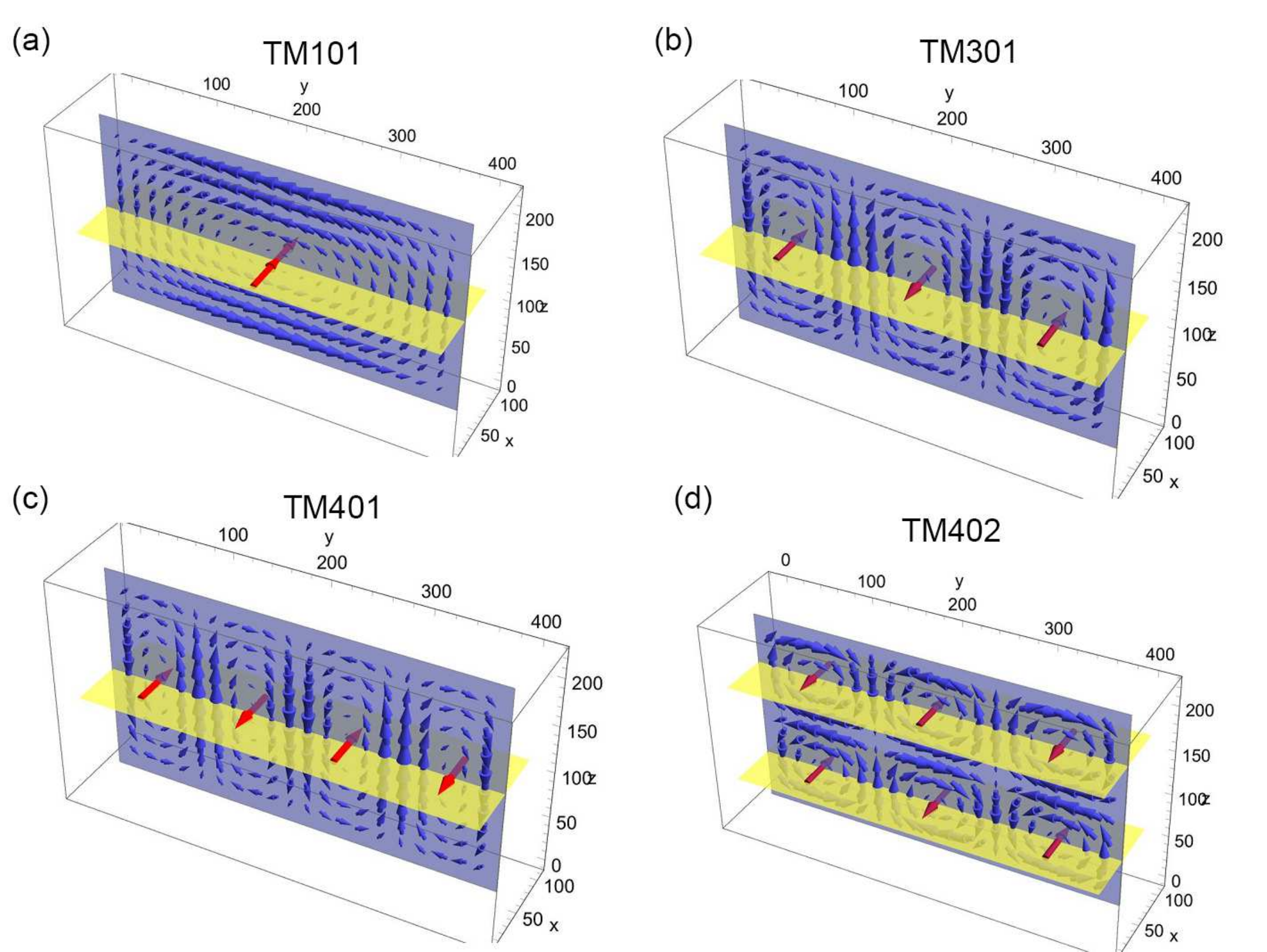}
\caption{Profile of the resonant cavity modes. Typical TM modes of the 
rectilinear cavity with quantum numbers (a) $n=1,m=0,l=1$, and (b) 
$n=3, m=0, l=1$, and (c) $n=4, m=0, l=1$, and (d) $n=3, m=0, l=2$. 
The blue arrows indicate electric vector field and red arrows magnetic 
vector field. The TM$_{101}$ and TM$_{301}$ modes replicate the field 
distribution of two magnetic dipole modes in Fig.3 in the main text.}
\end{figure}

Herein we present a theoretical interpretation of the near-field profiles of the 
hybrid Mie-Fabry-P\'{e}rot modes.  The optical resonances of a dielectric 
rectangular particle can be described in terms of induced standing waves 
inside a high impedance cavity.  Consider a homogeneous, isotropic dielectric 
rectangular resonator spanning $x=-W/2$ to $x=W/2$, $y=-L_y/2$ to 
$y=L_y/2$, and $z=-L_z/2$ to $z=L_z/2$. To decompose the electric and 
magnetic fields into standing wave cavity modes we begin with the vector 
Helmholtz equation, which can be obtained from the source-free Maxwell equations:
\begin{equation}
\nabla \times \nabla \times \{E,H\} - \omega^{2}\mu\varepsilon \{E,H\}=0.
\end{equation}
Solution of the vector Helmholtz equation (B1) can be obtained via the rectilinear 
generating function $\psi$
\begin{equation}
\nabla^{2}\psi-k^2\psi=0,
\end{equation}
where $k^{2}=\omega^{2}\mu\varepsilon$.
By separation of variables, the rectilinear generating function may be written as 
$\psi=X(x)Y(y)Z(z)$. Inserting this into the scalar Helmholtz equation (B2) and 
dividing by $X(x)Y(y)Z(z)$ yields:
\begin{equation}
\frac{1}{X}\frac{\partial^2{X}}{\partial{x^2}}+\frac{1}{Y}\frac{\partial^2{Y}}{\partial{y^2}}+\frac{1}{Z}\frac{\partial^2{Z}}{\partial{z^2}}=-k^2,
\end{equation}
from which we deduce
\begin{equation}
\frac{1}{X}\frac{\partial^2{X}}{\partial{x^2}}+k_x^{2}X = 0,\frac{1}{Y}\frac{\partial^2{Y}}{\partial{y^2}}+k_y^{2}Y = 0,\frac{1}{Z}\frac{\partial^2{Z}}{\partial{z^2}}+k_z^{2}Z = 0,
\end{equation}
with $k=k_x^{2}+k_y^{2}+k_z^{2}$. The general solution of equation (B3) can be written in the following form：
\begin{align}
X &= X_e\cos(k_xx)+X_o\sin(k_xx), \nonumber \\
Y &= Y_e\cos(k_yy)+Y_o\sin(k_yy), \nonumber \\
Z &= Z_e\cos(k_zz)+Z_o\sin(k_zz),
\end{align}
where the corresponding amplitudes are found from the corresponding boundary 
conditions. For high refractive index particles, due to their high impedance for 
the wave inside the cavity, perfect magnetic conductors (PMC) are typically 
used as approximate boundary conditions \cite{yee1965natural,okaya1962dielectric,mongia1997theoretical}.  
PMC boundary conditions are dual to the perfect electric conductor (PEC) conditions 
used for metallic cavities. Using PMC boundary conditions, i.e. 
$B_{\parallel}=E_{\perp}=0$, we can deduce the following electric and 
magnetic field profiles of the cavity modes:
\begin{align}
& \begin{pmatrix}
E_x\\
E_y\\
E_z
\end{pmatrix} 
= 
\begin{pmatrix}
A\sin(k_xx)\cos(k_yy)\cos(k_zz)\\
B\cos(k_xx)\sin(k_yy)\cos(k_zz)\\
C\cos(k_xx)\cos(k_yy)\sin(k_zz)
\end{pmatrix}, \nonumber \\
& \begin{pmatrix}
B_x\\
B_y\\
B_z
\end{pmatrix} 
= \frac{i}{\omega}\begin{pmatrix}
(Ck_y-Bk_z)\cos(k_xx)\sin(k_yy)\sin(k_zz)\\
(Ak_z-Ck_x)\sin(k_xx)\cos(k_yy)\sin(k_zz)\\
(Bk_x-Ak_y)\sin(k_xx)\sin(k_yy)\cos(k_zz)
\end{pmatrix}
\end{align}

Note that magnetic field satisfies the equation $\nabla \cdot B=0$. 
The coefficients $A$, $B$, $C$ are subject to the condition $\nabla\cdot E=0$, 
which leads to the condition  $Ak_x+Bk_y+Ck_z=0$. The boundary conditions 
determine the eigenfrequency of the cavity modes as:
\begin{align}
f &=\frac{\omega}{2\pi}=\frac{ck}{2\pi\sqrt{\varepsilon\mu}}=\frac{c}{2\pi\sqrt{\varepsilon\mu}}\sqrt{k_x^{2}+k_y^{2}+k_z^{2}} \nonumber \\
  & =\frac{c}{2\pi\sqrt{\varepsilon\mu}}\sqrt{(\frac{n\pi}{W})^2+(\frac{m\pi}{L_y})^2+(\frac{l\pi}{L_z})^2},\nonumber \\ 
f_{nml} &= \frac{c}{2\sqrt{\varepsilon\mu}}\sqrt{(\frac{n}{W})^2+(\frac{m}{L_y})^2+(\frac{l}{L_z})^2},
\end{align}
with $k_x = \frac{n\pi}{W},k_y = \frac{m\pi}{L_y},k_x = \frac{l\pi}{L_z}$. 
It should be noted that Eq. B7 holds both for dielectric and metallic cavities 
because of the duality of PEC and PMC conditions, whereas the electric and 
magnetic fields obtained in Eq. B6 for dielectric resonators are distinct from 
those for metallic cavities \cite{balanis2012advanced}. To relate this mode 
analysis to the scattering problem, we fix the direction of propagation along 
$z$-axis. For TM modes $B_z=0$, which requires that $Bk_x-Ak_y=0$, or 
$B=\frac{k_y}{k_x}A$ and $C=-\frac{A}{k_z}(\frac{k_y^2}{k_x}+k_x)$. 
This yields the $E$- and $B$-fields for TM$_{nml}$ modes:
\begin{widetext}
\begin{align}
\begin{pmatrix}
E_x\\
E_y\\
E_z
\end{pmatrix} 
= A\begin{pmatrix}
\sin(\frac{n\pi}{W}x)\cos(\frac{m\pi}{L_y}y)\cos(\frac{l\pi}{L_z}z)\\ 
\frac{mW}{nL_y}\cos(\frac{n\pi}{W}x)\sin(\frac{m\pi}{L_y}y)\cos(\frac{l\pi}{L_z}z)\\ 
-\frac{L_z(n^2L_y^2+m^2W^2)}{nlWL_y^2}\cos(\frac{n\pi}{W}x)\cos(\frac{m\pi}{L_y}y)\sin(\frac{l\pi}{L_z}z)
\end{pmatrix},\nonumber \\
\begin{pmatrix}
B_x\\
B_y\\
B_z
\end{pmatrix} 
= \frac{iA}{\omega}\begin{pmatrix}
-(\frac{mL_z(n^2L_y^2+m^2W^2)}{nlWL_y^3}+\frac{mlW}{nL_yL_z})\cos(\frac{n\pi}{W}x)\sin(\frac{m\pi}{L_y}y)\sin(\frac{l\pi}{L_z}z)\\ 
(\frac{L_z(n^2L_y^2+m^2W^2)}{lW^2L_y^2}+\frac{l}{L_z})\sin(\frac{n\pi}{W}x)\cos(\frac{m\pi}{L_y}y)\sin(\frac{l\pi}{L_z}z)\\ 
0
\end{pmatrix}
\end{align}
\end{widetext}

The cavity modes TM$_{101}$ and TM$_{301}$ [see Fig. 7(a,b)] replicate 
the electromagnetic field structure of two magnetic dipolar resonances in Fig. 3 
(see main text). One might also construct higher order magnetic dipole modes 
profiles for larger values of $n>1$ and $l>1$ [see Fig. 7(c,d)] and corresponding 
scattering resonant modes in Fig. 8 and Fig. 9. We would like to emphasize 
that this theoretical treatment is based on the approximate PMC boundary conditions 
which is only applicable to high-permittivity structures.  There is no exact closed-form 
expression available for the resonant frequencies or field distributions of such 
dielectric resonators, but we have provided an approximate solution to extract 
the essential modal behavior seen in simulations, as discussed in the main text.

\section{Radiation of MD$_{2}$ modes}\label{Appendix:C}
It is worth noting that for the conventional multipole decomposition, the MD mode
is usually defined as only one magnetic dipole positioned in the center of the particle 
(equation A4). However, here we can observe two separate magnetic dipoles 
in the near-field distributions of the MD$_2$ mode (Fig. 3c). Usually two dipoles cannot
be equivalently replaced by one dipole because the spatial distance between 
the two dipoles can lead to an additional phase term in the far-field response. 
However, in the following, we will show that, Eq. A4 and 
conventional multipole decomposition are still valid for MD$_2$ mode and can clearly reveal
its underlying physics.

First we consider two separate magnetic dipole \textbf{m$_1$} and \textbf{m$_2$}
at the MD$_2$ resonance with a spatial distance 2$d$, as shown in FIG. 8. 

\begin{figure}[h]
\includegraphics[width = 4.3cm]{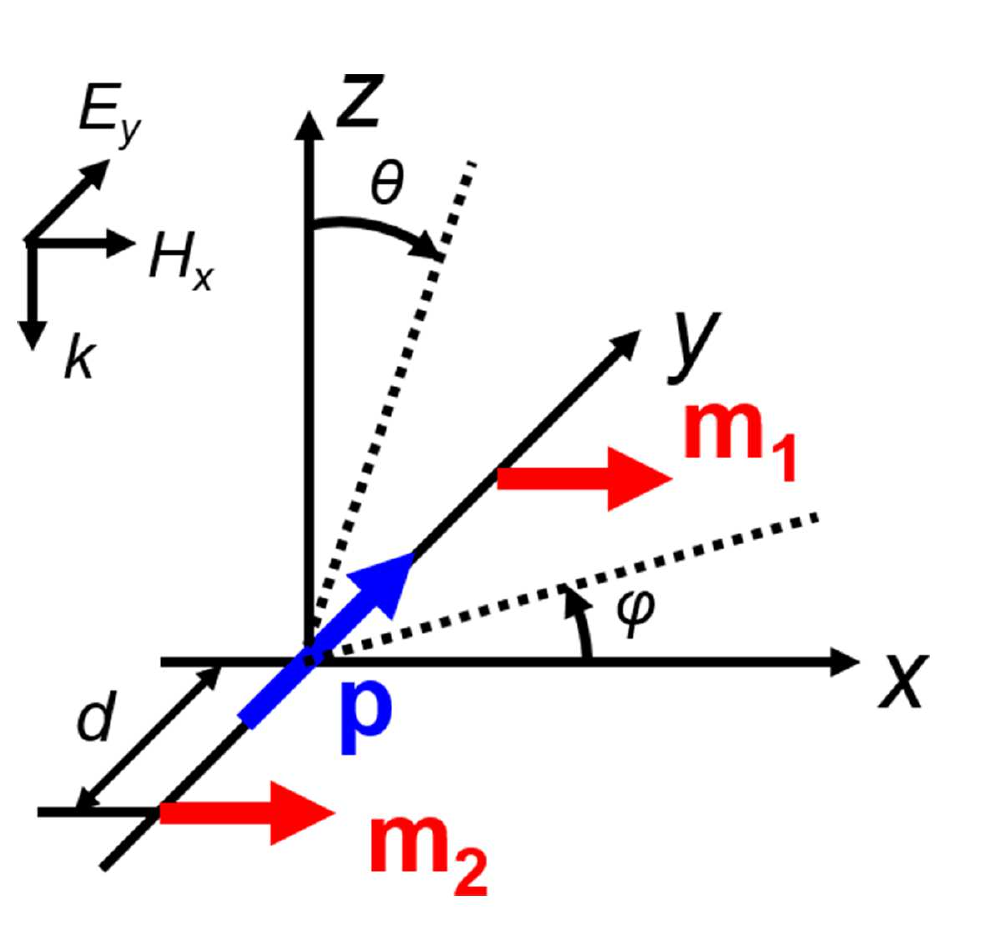}
\caption{Equivalent model for second magnetic resonance.}
\end{figure}

Given the axial symmetry of the structure, we can assume that these two magnetic 
dipoles are identical to each other with \textbf{m$_j$} $= \frac{1}{2}$ 
\textbf{m$_{\textnormal{MD2}}$} ($j = 1, 2$) , 
where \textbf{m$_{\textnormal{MD2}}$} is the total magnetic dipole moment 
that we can obtain through the multipole expansion. We note \textbf{y$_j$} 
the position vectors of the two magnetic dipoles and thus we can write the 
electric field \textbf{E$_{\textnormal{m}}$}  produced in the far-field by 
these two magnetic dipoles as:

\begin{equation}
\mathbf{E_\textnormal{m}}(\mathbf{r})= \sum_{j}\frac{k_0^2}{4\pi\epsilon_0rc}e^{ik|\mathbf{r-y_\textit{j}}|}(\mathbf{m_\textit{j}}\times\mathbf{n})
\end{equation}

At far-field limit where $r \gg d$ we can have:

\begin{align}
|\mathbf{r-y_\textit{j}}| - r &= \sqrt{x^2 + (y\mp d)^2 + z^2} - r \nonumber \\
& \approx r \Big( \sqrt{1 \mp \frac{2yd}{r^2}} - 1 \Big ) \nonumber \\
& \approx \mp d \Big( \frac{y}{r} \Big ) \approx \mp dsin\theta sin\varphi .
\end{align}

Then we can derive \textbf{E$_\textnormal{m}$} as follows:
\begin{align}
 \mathbf{E_\textnormal{m}} &= \frac{k_0^2}{4\pi\epsilon_0c}|\mathbf{m_\textnormal{MD2}}| \frac{e^{ikr}}{r} 
cos(kdsin\theta sin\varphi) \nonumber \\ 
& \cdot (-sin\varphi \hat{ \boldsymbol{\theta}} + cos\theta cos\varphi \hat{\boldsymbol {\varphi}}).
\end{align}

\noindent with $ \hat{ \boldsymbol{\theta}} $ and $\hat{\boldsymbol {\varphi}}$ 
the unit vectors of the spherical basis. In the above equation, one can clearly 
see the additional term  $cos(kdsin\theta sin\varphi)$ contributed by the spatial distance 
and how it influences the far-field response. However, this additional term will not have an 
impact on the total scattered power $P_\textnormal{m}$ contributed by the two magnetic dipoles, 
which can be determined by the following expression:

\begin{align}
P_m &= \int_{\Omega} d P_m d \Omega 
= \frac{1}{2Z_0} \int_{0}^{\pi} \int_{0}^{2\pi} |\mathbf{E_\textnormal{m}}|^2  r^2 sin\theta d \theta d\varphi \nonumber \\
& = \frac{Z_0 k^4}{12\pi} |\mathbf{m_\textnormal{MD2}}|^2.
\end{align}

Equation (C4) shows that the power contribution  $P_\textnormal{m}$ 
of two separate identical magnetic dipoles is only determined by 
their total magnetic dipole moment other than their relative positions. 
In our paper, we decompose the far-field scattering cross section into multipolar series, 
which is only related to the power contribution of each multipoles. Therefore, 
the second peak in the MD curve represents a local maximum contribution 
from the MD modes to the total scattering power, proving the existence of 
MD$_2$ mode which consists of two magnetic dipoles. 

Next, we consider the interference of the MD$_2$ and ED mode. As shown in FIG. 8, 
there is also an induced electric dipole \textbf{p} oscillating along $y$ direction. 
One can write the total electric field $\mathbf{E_\textnormal{pm}}$ induced 
by the three dipoles as: 

\begin{align}
\mathbf{E_\textnormal{pm}} (\mathbf{r}) &= \frac{k_0^2}{4 \pi \epsilon_{0} r} e^{ikr}
\Big[ |\mathbf{p}| (cos\theta sin\varphi \hat{\boldsymbol{\theta}} - cos\varphi \hat{\boldsymbol{\varphi}}) 
\nonumber \\ & + 2 \frac{\mathbf{|m_\textit{j}|}}{c} cos(kdsin\theta sin\varphi) (-sin\varphi \hat{ \boldsymbol{\theta}} + cos\theta cos\varphi \hat{\boldsymbol {\varphi}})
\Big].
\end{align}

Given the incident light is along $-z$ direction in our study, the backward and 
forward radar cross sections of the nanobar can be defined as:

\begin{align}
\sigma_{back} & = \lim_{r\to\infty}4\pi r^2 \frac{\mathbf{|E_\textnormal{pm}}(\theta = 0, \varphi =0)|^2}{|\mathbf{E_\textnormal{inc}}|^2} \nonumber \\
& = \frac{k^4}{4\pi\epsilon_0 |\mathbf{E}_\textnormal{inc}|^2} \left| {p_y - 2\frac{m_{xj}}{c}} \right| ^2,
\end{align}

\begin{align}
\sigma_{forward} & = \lim_{r\to\infty}4\pi r^2 \frac{\mathbf{|E_\textnormal{pm}}(\theta = \pi, \varphi =0)|^2}{|\mathbf{E_\textnormal{inc}}|^2} \nonumber \\
& = \frac{k^4}{4\pi\epsilon_0 |\mathbf{E}_\textnormal{inc}|^2} \left| {p_y + 2\frac{m_{xj}}{c}} \right| ^2,
\end{align}

\noindent with $|\mathbf{E_\textnormal{inc}}|$  is the amplitude of the incident electric field, 
$|p_y|$ and $|m_{xj}|$ are the amplitudes of the induced electric and magnetic dipole moments. 
Therefore, suppressed backscattering and maximum forward scattering occur 
if the following condition:

\begin{equation}
p_y = \frac{2}{c}m_{xj} = \frac{1}{c}m_{\textnormal{MD2}},
\end{equation}

\noindent is satisfied. Equation (C8) clearly reveals that, for a system consisting of ED and MD$_2$ modes,
unidirectional forward scattering can only appear when the electric dipole moment $p$ 
is equal to the total dipole moment $m_\textnormal{MD2}$ of the two magnetic dipoles $m_j$. 
When there is only one fundamental magnetic dipole, equation (C8) becomes 
$p_y = \frac{1}{c}m_x$, which is the well-known first 
Kerker condition \cite{kerker1983electromagnetic}.

\section{Near field distributions of higher-order hybrid modes}\label{Appendix:D}

As predicted by the theory (Fig. 7) and demonstrated by the numerical simulations 
(Fig. 4), we expect to find higher-order hybrid modes accompanied by higher-order 
multipoles and cavity modes with increasing geometric parameters. 
Here we show their near-field distributions.

\begin{figure}[t]
\includegraphics[width = 8.6cm]{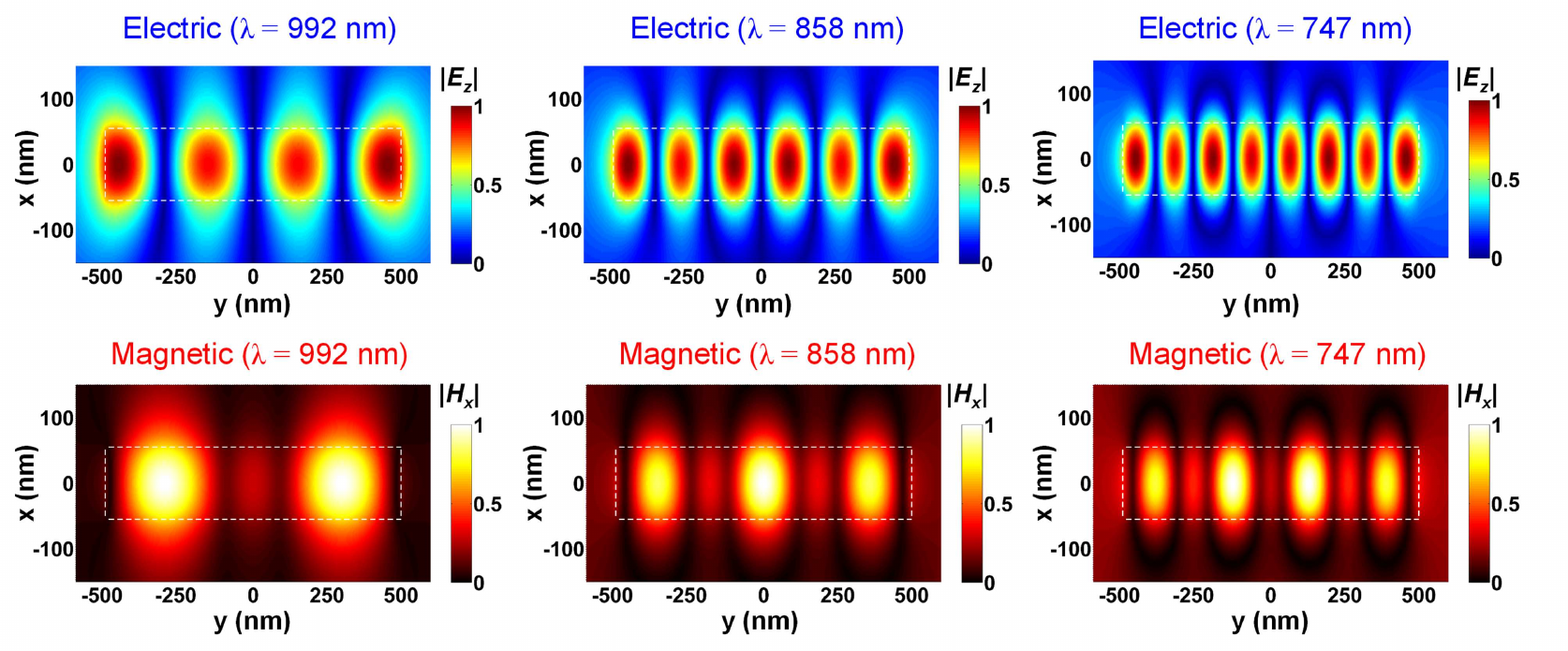}
\caption{Near-field distributions of electric and magnetic fields of the nanobar with 
$L_y$ = 1000 nm, $W$ = 110 nm, $L_z$ = 220 nm.}
\includegraphics[width = 8.6cm]{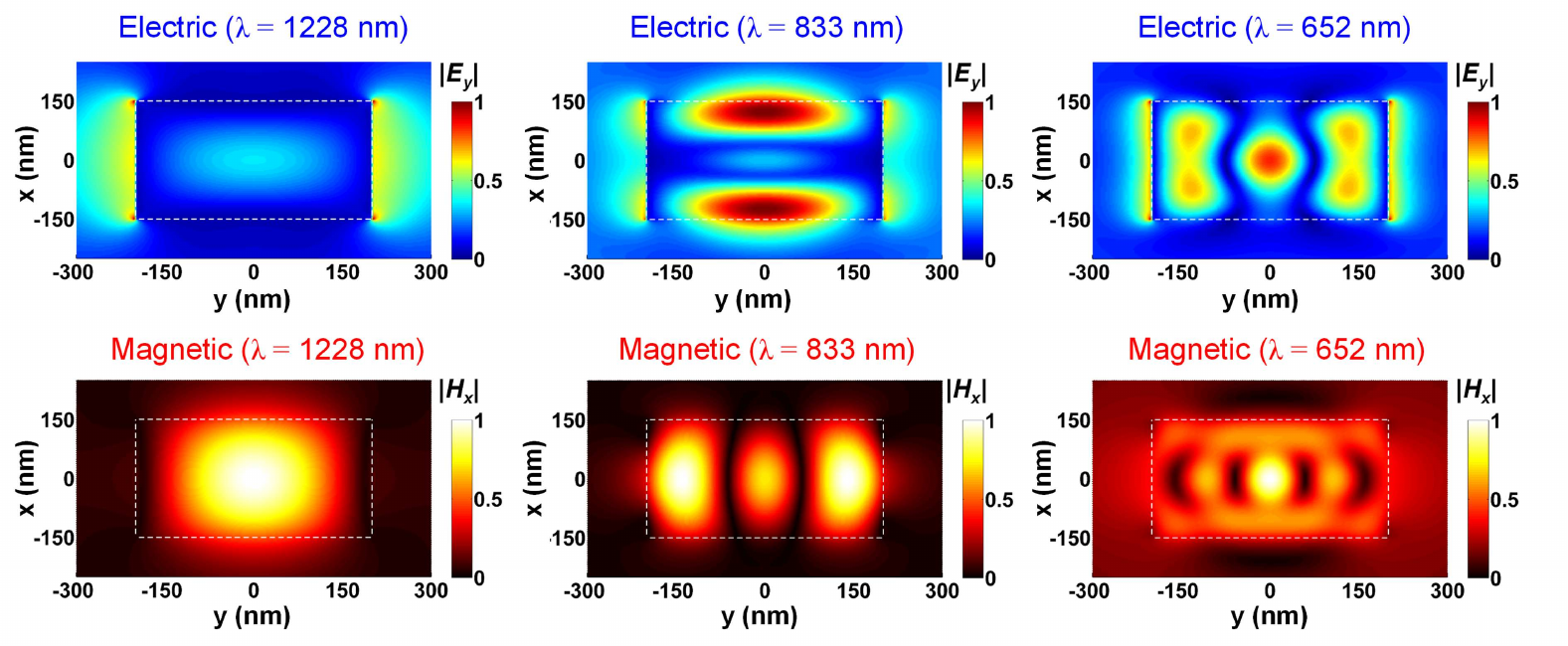}
\caption{Near-field distributions of electric and magnetic fields of the anisotropic nanobar with $W$ = 300 nm, $L_y$ = 400 nm, $L_z$ = 220 nm.}
\end{figure}

For increased length $L_y$ = 1000 nm, to clearly illustrate the ``higher-mode" 
magnetic dipolar responses and the associated higher-order cavity modes, 
here we plot field components $E_z$ and $H_x$. It can be readily seen that 
the three peaks ($\lambda$ = 992 nm, 858 nm and 747 nm) in the scattering spectrum 
(see Fig. 4(c)) correspond to the existences of MD$_2$, MD$_3$ and MD$_4$ mode, 
respectively.

For increased width $W$ = 300 nm, fundamental ED and MD modes can be clearly seen 
at $\lambda$ = 1228 nm, while at $\lambda$ = 833 nm, an ED$_2$ mode accompanied 
by a standing wave pattern (3 anti-nodes) in $x$ direction can be observed. A MD$_2$ 
mode can also be seen at this wavelength. For shorter wavelength $\lambda$ = 652 nm, 
we observe complex and hybrid modal distributions while the higher-mode ED and MD 
responses could still be distinguished.

\begin{acknowledgments}
YY and MQ acknowledge financial support from the National Natural Science
Foundation of China (Grant No. 61425023, 61575177, 61275030, 61235007). AEM and
YSK were supported by the Australian Research Council. SVK acknowledges support
by the U.S. National Science Foundation under Grant No. 1515343. The numerical simulation and 
experimental investigation of the metasurfaces in microwave frequency range were
supported by the Russian Science Foundation (Project No. 14-12-00897). 
PK acknowledges the scholarship of the President of Russian Federation.
\end{acknowledgments}

\bibliography{M1}

\end{document}